\documentclass[12pt]{iopart}

\usepackage{iopams}
\usepackage{graphicx}

\begin{document}

\title[Controlling the quantum computational speed ]
{\bf Controlling the quantum computational speed}

\author[Metwally et al]{N. Metwally, M. Abdel-Aty\footnote[1]{E-mail:
abdelatyquantum@gmail.com} and M. Sebawe Abdalla }
\address{Mathematics Department, College of Science, Bahrain University,
32038 Kingdom of Bahrain
\\
Mathematics Department, Faculty of Science, Sohag University,
82524 Sohag, Egypt
\\
Mathematics Department, College of Science, King Saud University,
Riyadh 11451, Saudi Arabia }

\begin{abstract}

The speed of quantum computation is investigated through the time
evolution of the speed of the orthogonality. The external field
components for classical treatment beside the detuning and the
coupling parameters for quantum treatment play important roles on
the computational speed. It has been shown that the number of
photons has no significant effect on the speed of computation.
However, it is very sensitive to the variation in both detuning
and the interaction coupling parameters.

\end{abstract}

\[
\]
%{Submitted to:} {\it J. Phys. A}

\maketitle

\section{Introduction}

Recently, there are a great interest for developing the computer
device. This is mainly due to the complications of the problems
which we are usually facing and urgently need to find their
solutions \cite{gor05,ben98,bri05}. To overcome these problems we
have to find a computer device with high capacity and speed or we
have to develop a new kind of computer has such properties. This
in fact means that we need one has a large memory, an adequate
processor and a large hard disk. This stimulated and encouraged
the researcher to seek for another kind of computer different to
that of the classical one, that is the quantum computer. However,
in the quantum information and more precisely in the quantum
computer there is an important question would be raised, what is
the speed of sending information from nod to the other one that to
reach the final output. Since the information is coded in a
density operator, therefore we ask how fast the density operator
will change its orthogonality. In other words, we search for a
minimum time needed for a quantum system to pass from one
orthogonal state to another \cite{Norman}. To perform this task we
need qubit contains the information to evolve through a unitary
operator where the carrier transforms it from one nod to the
other.

For entangled qubit pair, one can see the operators cause a decay
of entanglement \cite{Eberly,Ban,Mban}. Further, it has been shown
that the classical noise leads to what is called entanglement
sudden death \cite{Eb}. Moreover, the time-dependent interaction
of a single qubit with a field can also produce such a phenomena
as well as along-lived entanglement\cite{Aty}. This means that
there are different factors would be involved and affect the
transmission speed and consequently the information. Recently the
efforts in quantum information research are directed towards
improving the performance of single qubit interaction. Also,
evolution speed (maximum transition rate between orthogonal state)
and the time evolution of some models has been discussed in Ref.
\cite{saw04}.

The main purpose for the present communication is to consider the
interaction between a single qubit and an external field for the
classical treatment and the interaction between a single qubit and
cavity field for the quantum treatment. This is to shed some light
on the general behavior of the interaction process and its
relationship with the speed of the computation \cite{Norman}
(maximum number of orthogonal states that the system can pass
through per unit time), speed of orthogonality \cite{yun06}
(minimum time for a quantum state $|\psi_i\rangle$ to evolve into
orthogonal state $|\psi_f\rangle$ where
$\langle\psi_i|\psi_f\rangle=0$) and speed of evolution
\cite{fal06} (maximum transition rate between orthogonal state).

The paper is organized as follows. In Sec. $2$, we consider the
classical treatment, where we calculate the general form of the
time evaluation of the density operator. The quantum interaction
of the qubit will be considered in Sec. $3$, where we introduce
the unitary operator in an adequate form. Also, we obtain the
final state by means of the Bloch vectors. Also we study the
effect of the field parameters on the speed of the quantum
computation. Finally, our conclusion is given in Sec $4.$

\section{Classical treatment of Qubit}

Let us start out with a short reminder on a general form of the
density operator of a qubit with the aid of analogs of Pauli's
spin vector operator $\overrightarrow\sigma$. This row vector
refer to the three dimensional vector \cite{Englert}
\begin{equation}
\overrightarrow{\sigma }=\sum_{\alpha
=x,y,z}{\overrightarrow{\sigma _{\alpha }}{e_{\alpha }^{\downarrow
}}}=(\sigma _{x},\sigma _{y},\sigma _{z})\left(
\begin{array}{c}
\overrightarrow e_{x}\\
\overrightarrow e_{y}\\
\overrightarrow e_{z}
\end{array}%
\right),
\end{equation}%
where $\overrightarrow e_{x,y,z}$ are orthonormal vectors of the
three coordinate axes to which the components $\sigma_{i}$ are
referred and $\sigma_{i}$ are Pauli matrices satisfying the
commutation relation $[\sigma _{i},\sigma _{j}]=2i\sigma _{k},$
where $i,j,k$ form an even permutation of $x,y,z$. In this case,
the density operator can be represented in the following form
\cite{fal06}
\begin{eqnarray}
\rho_a
&=&\frac{1}{2}(1+\overrightarrow{s}\otimes{\sigma^{\downarrow }})
\nonumber
\\
&=&\frac{1}{2}(1+s_{x}\sigma_{x}+s_{y}\sigma_{y}+s_{z}\sigma _{z})
\nonumber
\\
&=& \frac{1}{2}\left(
\begin{array}{c}
1+s_{z} \\
s_{x}-is_{y}%
\end{array}%
\begin{array}{c}
s_{x}+is_{y} \\
1-s_{z}%
\end{array}%
\right). \label{eq1}
\end{eqnarray}
where $\overrightarrow{s}=\langle {\overrightarrow{\sigma }\rangle
}$.

 The effective Hamiltonian for one qubit can be defined as
\cite{mot06}
\begin{equation}
H=\alpha_x\sigma _{x}+\alpha_y\sigma _{y}+\alpha_z\sigma _{z},
\end{equation}%
where $\alpha _{i}$ are the external field components. The unitary
evolution operator can be obtained from the Hamiltonian, thus,
\begin{equation}
\mathcal{U}=\sum_{i}\exp\{-i\alpha _{i}t\sigma_{i}  \},
\label{uni}
\end{equation}%
Using the density operator (2), Alice qubit can be transformed as
\begin{equation}
\rho _{a}\rightarrow \tilde{\rho}_{a}=\frac{1}{2}(1+\tilde{%
\mathord{\buildrel{\lower3pt\hbox{$\scriptscriptstyle\rightarrow$}}\over s}}%
\otimes{\sigma
^{\raisebox{2pt}[\height]{$\scriptstyle\downarrow$}}}),
\end{equation}%
where the component of
$\tilde{\mathord{\buildrel{\lower3pt\hbox{$\scriptscriptstyle\rightarrow$}}\over
s}}$ are given by
\begin{eqnarray}
\tilde{s_{x}} &=&\frac{1}{3}\bigl[s_{x}(1+\cos (2t\alpha
_{2})+\cos (2t\alpha _{3}))-s_{y}\sin (2t\alpha _{3})+s_{z}\sin
(2t\alpha _{2})\bigr]
\nonumber \\
\tilde{s_{y}} &=&\frac{1}{3}\bigl[s_{y}(1+\cos (2t\alpha
_{1})+\cos (2t\alpha _{3}))+s_{x}\sin (2t\alpha _{3})-s_{z}\sin
(2t\alpha _{1})\bigr],
\nonumber \\
\tilde{s_{z}} &=&\frac{1}{3}\bigl[s_{z}(1+\cos (2t\alpha
_{1})+\cos (2t\alpha _{2}))-s_{x}\sin (2t\alpha _{2})+s_{y}\sin
(2t\alpha _{1})\bigr].
\nonumber \\
&&
\end{eqnarray}%
Having obtained the above analytical expressions for $\tilde{s_{x}},\tilde{%
s_{y}}$ and $\tilde{s_{z}}$, we are therefore in position to
investigate the speed of the orthogonality and hence the speed of
computation. To clarify our idea let us assume that the user Alice
has encoded some information in her qubit which is defined by
\begin{equation}
\rho _{a}=\frac{1}{2}(1+s_{x}\sigma _{x}).  \label{in:eq1}
\end{equation}%
Using the time evolution of the unitary operator, (\ref{uni}) one
can transform $\rho _{a}$ into $\tilde{\rho}_{a}$ from which the
new Bloch vectors take the form,
\begin{equation}
\tilde{s_{x}}=\frac{s_{x}}{3}(1+\cos 2t\alpha _{2}+\cos 2t\alpha
_{3}),\quad
\tilde{s_{y}}=\frac{s_{x}}{3}\sin 2t\alpha _{3},\quad \tilde{s_{z}}=-\frac{%
s_{x}}{3}\sin 2t\alpha _{2}.  \label{eq:FB}
\end{equation}%
Let us assume that Alice has prepared her qubit such as $s_{x}=1$ and $%
s_{y}=s_{z}=0$. Then the eigenvectors of the (\ref{in:eq1}) state
can be written as
\begin{equation}
v_{1}=[-1,1],\quad \mbox{and\quad}v_{2}=[1,-1].
\end{equation}%
Thus, it will be easy to get the eigenvectors for the final state $\tilde{%
\rho}_{a}$, which is described by Bloch vectors (\ref{eq:FB}).
After some algebraic calculations, we can explicitly write $u_{i}$
as
\begin{equation}
u_{1}=u_{2}=\Gamma \biggl\{\lbrack \sin ^{2}2t\alpha _{2}-\left(
3+\cos 2t\alpha _{2}+\cos 2t\alpha _{3}+\cos 2t\alpha _{2}\cos
2t\alpha _{3}\right) ]\biggr\},
\end{equation}%
where $\Gamma =(1+\cos 2t\alpha _{2}+\cos 2t\alpha _{3}-i\sin
2t\alpha _{3})^{-1}$. In order to facilitate our discussion let us
define the scalar product of the vectors $u_{i}$ and $v_{i}$ such
as
\begin{equation}
Sp_{ij}=\left\langle v_{i}|u_{j}\right\rangle .
\end{equation}%
\begin{figure}[tbp]
\begin{center}
\includegraphics[width=18pc,height=10pc]{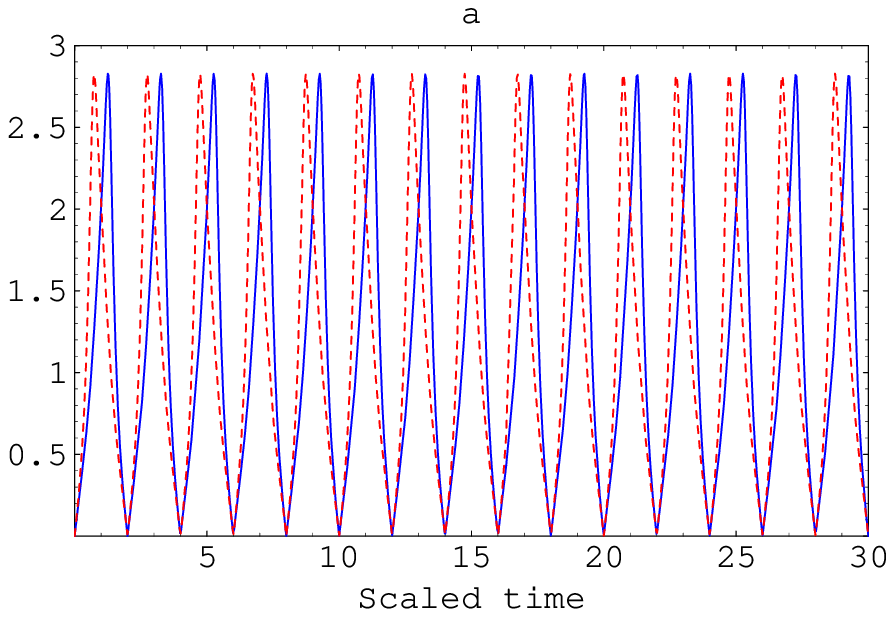} %
\includegraphics[width=18pc,height=10pc]{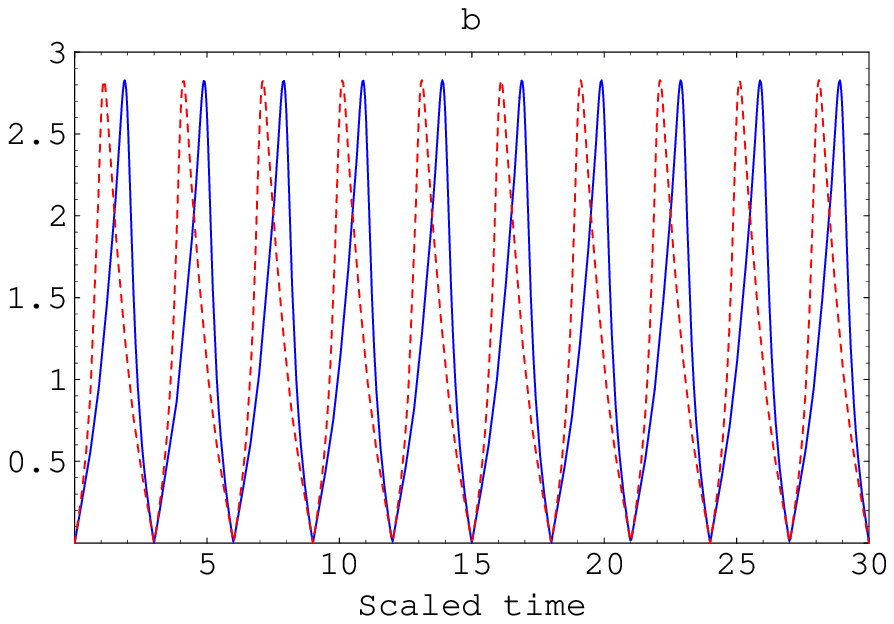} %
\includegraphics[width=18pc,height=10pc]{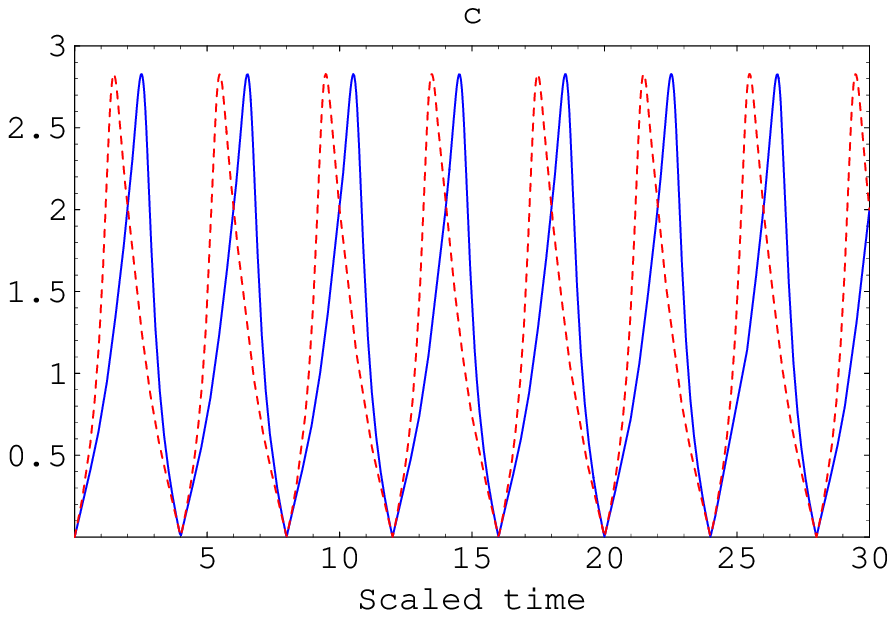} %
\includegraphics[width=18pc,height=10pc]{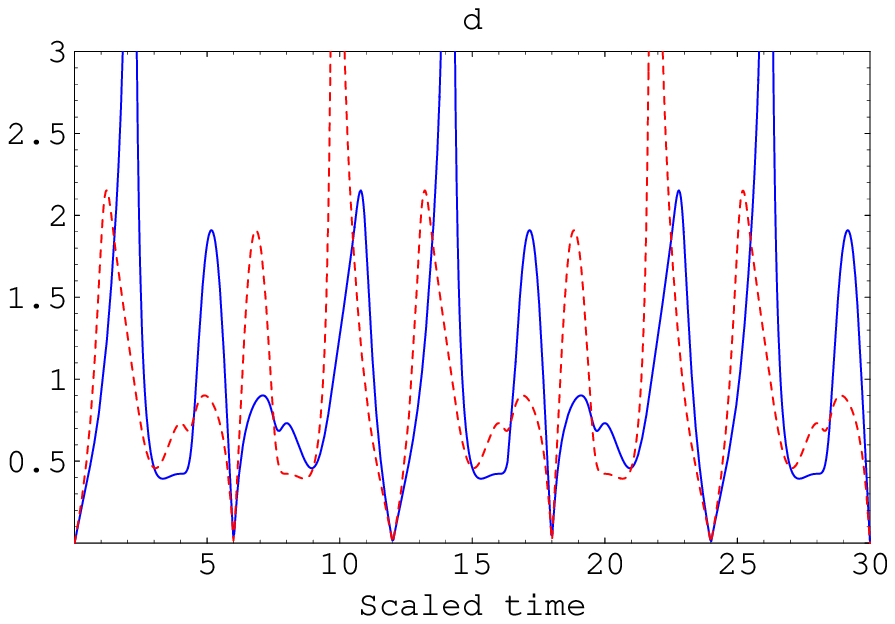}
\end{center}
\caption{The speed of orthogonality of qubit as a function of the
scaled time, where, the component $\left\langle
u_{i}|v_{1}\right\rangle $ is represented by the solid curve,
while $\left\langle u_{i}|v_{2}\right\rangle$ is represented by
the dotted-curve. The other parameters are $s_{x}=1$ and
$s_{y}=s_{z}=0$, (a) $\alpha _{1}=\alpha _{2}=\alpha
_{3}=\frac{\pi }{2}$, (b) $\alpha _{1}=\alpha _{2}=\alpha
_{3}=\frac{\pi }{3}$, (c) $\alpha _{1}=\alpha _{2}=\alpha
_{3}=\frac{\pi}{4}$ and (d) $ \alpha _{1}=\frac{\pi }{2}$,
$\alpha_{2}=\frac{\pi}{3}$ and $\alpha_{3}=\frac{\pi}{4}$. }
\end{figure}
It should be noted that in our calculations we have taken into
account all the possible products of $u_{i}$ and $v_{i}$. In
figure (1) we have plotted the amplitude values of $Sp_{ij}$
against the scaled time to display its behavior for different
values of the control parameter $\alpha _{i}.$ In figure $(1a)$ we
have considered the case in which $\alpha _{i}=\pi /2$ where one
can see both $\langle v_{1}|u_{j}\rangle $ and $\langle
v_{2}|u_{j}\rangle $ are coincides on the horizontal axis at
different period of time. However, when we change the value of the
parameters $\alpha _{i}$ such as $\alpha _{i}=\pi /3$ it is noted
that there is decreasing in the number of coincidences points
which refer to reduction in the computation speed, see figure
(1b). This phenomenon gets more pronounced for the case in which
$\alpha _{i}=\pi /4$, see figure (1c). Thus we may conclude that
as the value of the control parameters $\alpha _{i}$ increases as
the speed of the computation increases and vise versa. On the
other hand when we consider different values for the
control parameters $\alpha _{i}$ such that $\alpha _{1}=\frac{\pi }{2}%
,~\alpha _{2}=\frac{\pi }{3},~\alpha _{3}=\frac{\pi }{4}$, then
more decreasing can be seen in the computation speed. In the
meantime we can observe irregular fluctuations in both functions
$Sp_{1j}$ and $Sp_{2j}$ in addition to the intersection at
different points, see figure (1d). This is contrary to the
previous cases where regular oscillations can be realized in each
case.

\section{Quantum treatment of Qubit}

Now let us turn our attention to consider the quantum treatment of
the computation speed, taking into account the quantized field
interacting with a single qubit. In this case the Hamiltonian can
be written as,
\begin{equation}
\hat{H}_{int}=\lambda (\hat{a}^{\dagger }\sigma _{-}+\hat{a}\sigma _{+})+%
\frac{\Delta }{2}\sigma _{z},
\end{equation}%
where $\hat{a}^{\dagger }$ and $\hat{a}$ are the creation and
annihilation operators satisfy the commutation relation
$[\hat{a},\hat{a}^{\dagger }]=1.$ We denote by $\lambda $ the
coupling constant and $\Delta $ the detuning parameter while
$\sigma _{+}(\sigma _{-}),$ $\sigma _{z}$ are usual raising
(lowering) and inversion operators for the two-level atomic
system, satisfying $[\sigma _{z},\sigma _{\pm }]=\pm 2\sigma _{\pm
}$ and $[\sigma _{+},\sigma _{-}]=2\sigma _{z}$ . The
time-dependent density operator $\rho (t)$ is given by
\begin{figure}[tph]
\begin{center}
\includegraphics[width=18pc,height=10pc]{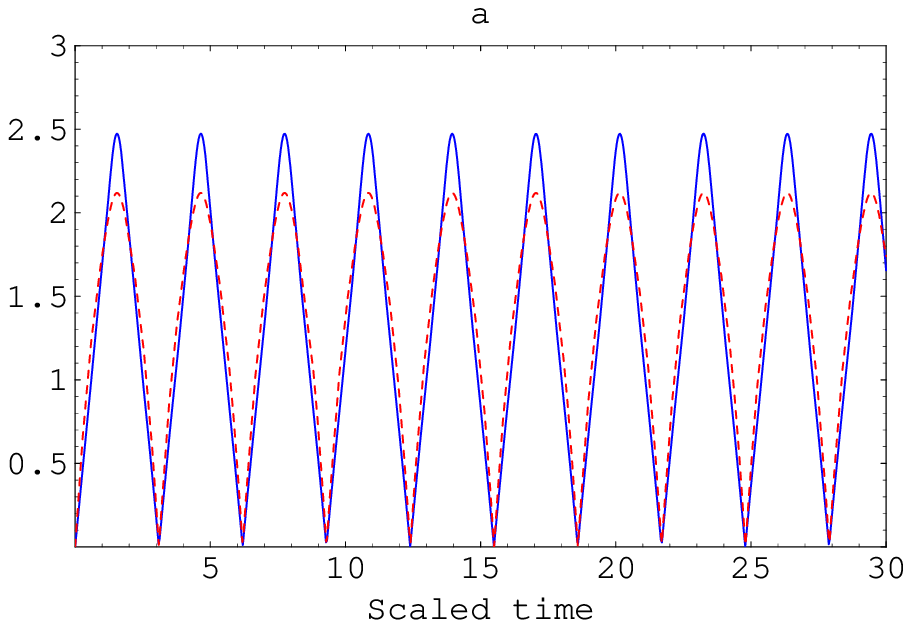} %
\includegraphics[width=18pc,height=10pc]{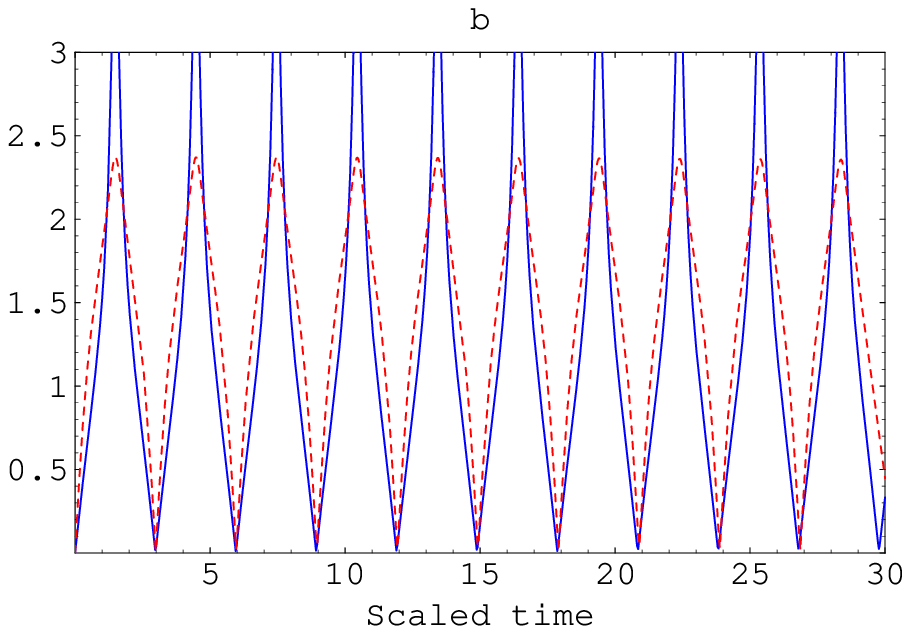} %
\includegraphics[width=18pc,height=10pc]{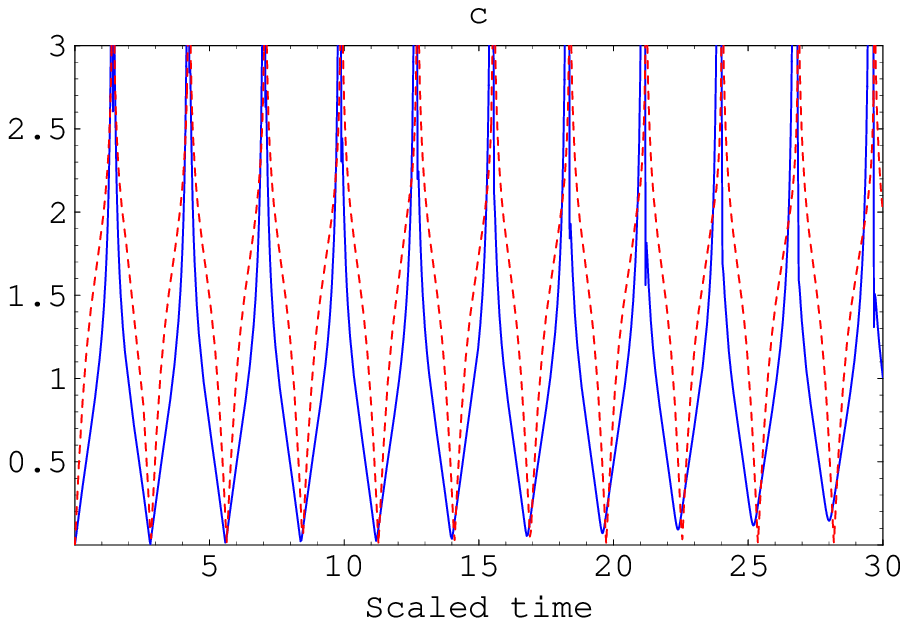} %
\includegraphics[width=18pc,height=10pc]{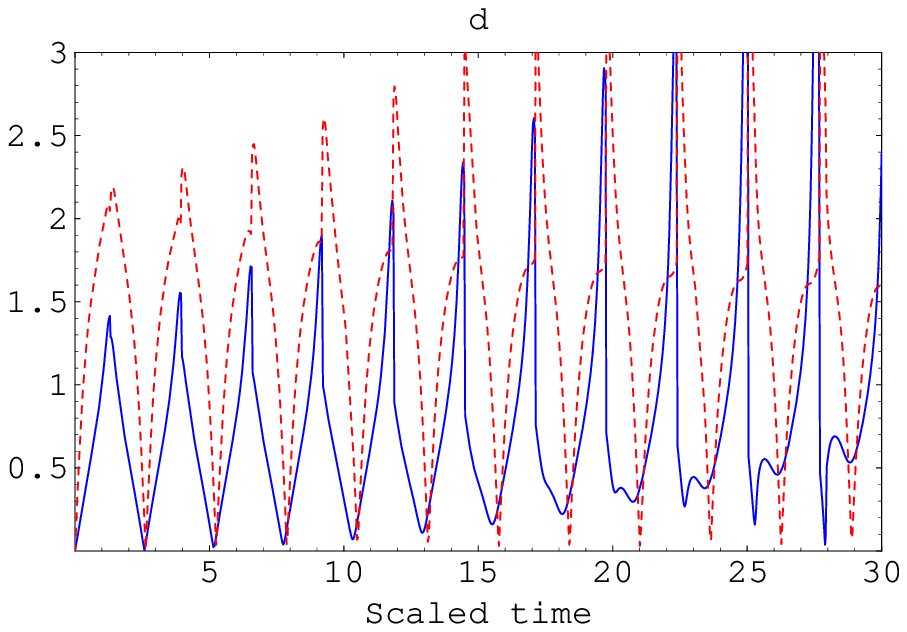}
\end{center}
\caption{The same as figure 1, but $n=10,~\Delta /\protect\gamma =2$ (a)~$%
\protect\eta =0.05$, (b) $\protect\eta =0.1$, (c)~ $\protect\eta
=0.15$ and (d)~$\protect\eta =0.2$. }
\end{figure}
\begin{equation}
\rho (t)=\mathcal{U}(t)\rho (0)\mathcal{U}^{\dagger }(t),
\label{in2}
\end{equation}%
where $\rho (0)=\rho _{a}(0)\otimes \rho _{f}(0)$ is the initial
state of the system. We assume that the initial state of the qubit
$\rho _{a}(0)$ is given by equation (7) while the field starts
from a Fock state. Under the above assumptions the unitary
evolution operator $\mathcal{U}(t)$ can be defined as
\begin{equation}  \label{un22}
\mathcal{U}(t)=u_{11}\bigl|0\bigr\rangle\bigl\langle0\bigr|+u_{12}\bigl|1%
\bigr\rangle\bigl\langle0\bigr|+u_{21}\bigl|0\bigr\rangle\bigl\langle1\bigr|%
+u_{22}\bigl|1\bigr\rangle\bigl\langle1\bigr|  \label{un21}
\end{equation}%
where
\[
u_{11}=C_{n+1}-i\frac{\Delta }{2}S_{n+1},\quad u_{12}=-i\eta
S_{n}a,\quad u_{21}=u_{12}^{\dagger },\quad u_{22}=u_{11}^{\dagger
},
\]%
and
\[
S_{n}=\frac{\sin (\mu _{n}\gamma t)}{\mu _{n}},\qquad C_{n}=\cos
(\mu _{n}\gamma t),\qquad \mu _{n}=\sqrt{\frac{\Delta
^{2}}{4\gamma ^{2}}+\eta ^{2}n}.
\]%
It should be noted that in the above equations, we have introduced
the
parameters $\eta $ and $\gamma $ to connect up the coupling parameter $%
\lambda $ such that $\lambda =\eta \gamma .$ This in fact would
enable us to
discuss the effect of the coupling parameter using $\eta $ instead of $%
\lambda $ regarding $\gamma $ as a dimensionless parameter. Using Eqs.(\ref%
{in2}) and (\ref{un22}) one can obtain the explicitly
time-dependent density operator $\rho (t)$ in Bloch vectors
representation, thus
\begin{eqnarray}
\tilde{S_{x}} &=&-i\eta S_{n+1}\frac{1+s_{z}}{2}\bigl[\sqrt{n+1}(C_{n+1}+%
\frac{\Delta }{2}S_{n+1})+(C_{n+1}-\frac{\Delta }{2}S_{n+1})\bigr]
\nonumber
\\
&&+\eta ^{2}\sqrt{n+1}\sqrt{n+2}S_{n}S_{n+1}s_{x}+i\eta \sqrt{n+1}\frac{%
1-s_{z}}{2}S_{n}C_{n+2}  \nonumber \\
&&+(C_{n+1}^{2}-\frac{\Delta ^{2}}{2}S_{n+1}^{2})s_{x}-\Delta
S_{n+1}C_{n+1}s_{y},  \nonumber \\
\tilde{S_{y}} &=&\eta S_{n+1}\frac{1+s_{z}}{2}\bigl[\sqrt{n+1}(C_{n+1}+\frac{%
\Delta }{2}S_{n+1})+(C_{n+1}-\frac{\Delta }{2}S_{n+1})\bigr]  \nonumber \\
&&-\eta ^{2}\sqrt{n+1}\sqrt{n+2}S_{n}S_{n+1}s_{y}+i\eta \Delta \sqrt{n+1}%
\frac{1-s_{z}}{2}S_{n}S_{n+2}  \nonumber \\
&&+\Delta S_{n+1}C_{n+1}s_{x}+(C_{n+2}^{2}-\frac{\Delta ^{2}}{2}%
S_{n+1}^{2})s_{y},  \nonumber \\
\tilde{S_{z}} &=&-i\frac{\eta }{2}\bigl[(1+\sqrt{n+1})C_{n+1}+\frac{\Delta }{%
2}(1-\sqrt{n+1})S_{n+1}\biggr]s_{x}S_{n+1}  \nonumber \\
&&+\frac{\eta }{2}\bigl[(1-\sqrt{n+1})C_{n+1}+\frac{\Delta }{2}(1+\sqrt{n+1}%
)S_{n+1}\biggr]s_{y}S_{n+1}  \nonumber \\
&&-(C_{n+1}^{2}+\frac{\Delta ^{2}}{4}S_{n+1})s_{z}+\eta ^{2}S_{n}^{2}[\frac{1%
}{2}-(n+\frac{1}{2})s_{z}]  \nonumber \\
&&-i\eta \frac{s_{x}-is_{y}}{2}\sqrt{n+1}S_{n}C_{n+1}.
\end{eqnarray}%
The parameters $s_{x},s_{y}$ and $s_{z}$ which appear in the right
hand side of equation (15), describe the initial state Bloch
vectors.

\begin{figure}[b]
\begin{center}
\includegraphics[width=18pc,height=10pc]{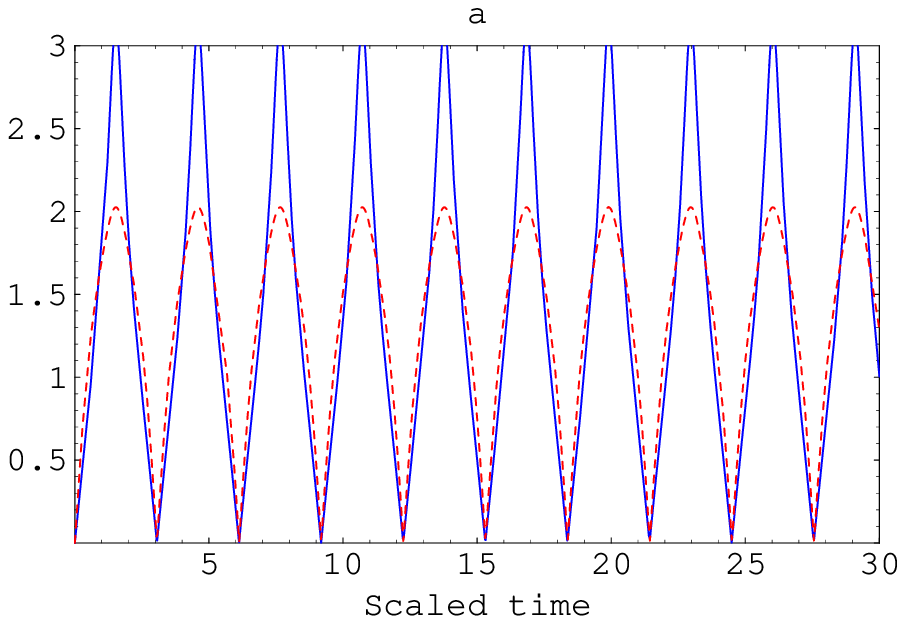} %
\includegraphics[width=18pc,height=10pc]{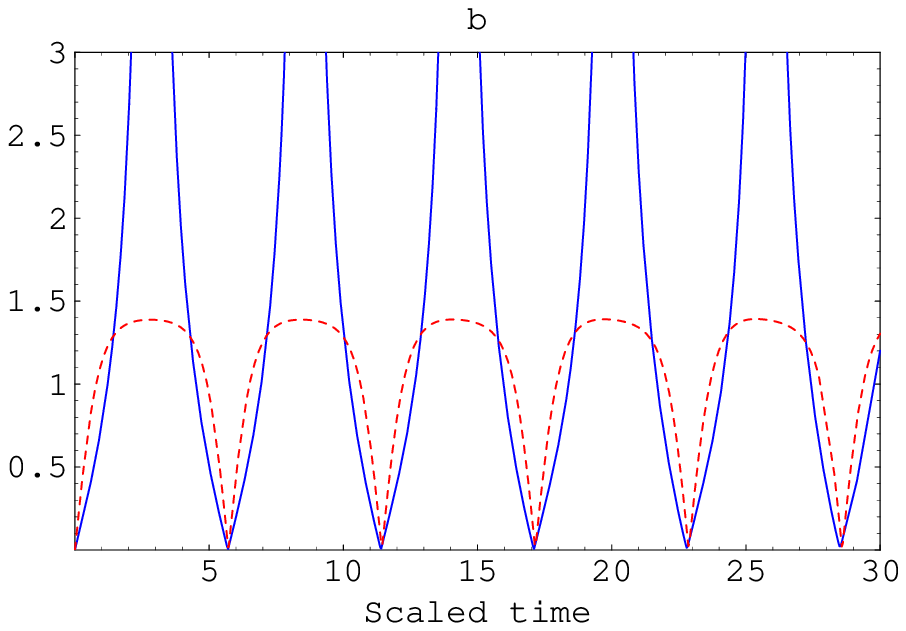}
\end{center}
\caption{The same as figure (2a), but $n=20$, (a)~$\Delta/\protect\gamma =2$%
, (b) $\Delta/\protect\gamma =1$. }
\end{figure}

In fact, these states are widely used in the quantum information
tasks. For example, one may consult a recent applications given in
Ref. \cite{Nielsen}. In the meantime it would be interesting to
employ the Hamiltonian given by equation (12) that to discuss the
speed of computation. This is extensively used in the field of
quantum information to describe the interaction between field and
qubit, particularly for investigating the loss of entanglement
after propagation in a quantum noisy channel. Furthermore, compare
with the classical treatment the interaction Hamiltonian contains
three different parameters to control the dynamics of the system,
$\Delta $ the detuning parameter, $\eta$ the coupling constant,
and the mean photon number $\langle a^{\dagger }a\rangle$
\cite{Brito}. These parameters are involved through the Rabi
frequency $\lambda _{n}$ as well as in the Bloch vectors
themselves.
This would give us a wide range of variety to discuss the variation in $%
Sp_{ij}$ resultant of change one of these parameters. To do so we
have numerically calculated the overlap between the initial and
the final states of $Sp_{ij}$. For example, to see the effect of
the coupling parameter $\eta $we have considered the number of
photons $n=10$, and the detuning parameter $\Delta/\gamma =2,$
while $\eta =0.05.$

In this case and from figure (2a) we can see nearly perfect
overlap between both of $Sp_{1j}$ and $Sp_{2j}$ as well as
coincidences at the horizontal line showing high speed. Increase
the value of $\eta$ such that $\eta =0.1$ leads to slight increase
in the speed of computation, beside increases in one of the
projectors value, see figure (2b). More increases in the coupling
parameter $\eta =0.15$ shows increasing in the speed of
computation but with less coincidences between the two projectors,
see figure (2c). More increasing in the coupling value $\eta =0.2$
leads to more decreasing in the speed of computation but with
regular increasing in both projectors value. This means that there
is a certain value (critical value) of the coupling parameter
where the speed of computation reaches its maximum and then starts
to slow down. To examine the effect of the mean photon number we
have considered the case in which $n=20,$ keeping the other
parameters unchange as in figure (2a). In this case we observe no
change in the speed of orthogonality and the behavior in general
is the same as before, however, there is increasing in the
amplitude for one of the projectors, see figure
(3a). However, if we decrease the value of the detuning parameter $%
\Delta/\gamma=1$ drastic change can be realized. For instance, we
can see decreasing in the number of the oscillations period,
increasing in one of the projector amplitude, in addition to
decrease in the speed of computation, see figure (3b). Thus we
come to conclusion if one increases the value of the detuning
parameter then the speed of the interaction
increases. This result is in agreement with that given by Montangero \cite%
{Simon}, where they investigated the dynamics of entanglement in
quantum computer with imperfections.

\section{Conclusion}

In the above sections of the present paper we have considered the
problem of speed computation in quantum information. The problem
has been handled from two different point of view; where we have
considered both of classical and quantum treatments. The main
concentration was on how to improve and control the computation's
speed in each case separately. For the classical treatment it has
been shown that the speed of computation is proportional with the
total value of the external field. However, for quantum treatment
we have seen that the speed of computation is sensitive to the
variation of the coupling parameter and the detuning parameter. In
the meantime we found the mean photon number does not play any
role with the speed of computation but it is just effect the
amplitude of the projectors. This in fact would turn our attention
to look for the atom-atom interaction to be discussed in a
forthcoming work.

 \textbf{Acknowledgements:}

One of us (M.S.A) is grateful for the financial support from the
project Math 2005/32 of the research center, College of Science,
King Saud University.

\end{document}